\documentclass[aps,prd,onecolumn,11pt,superscriptaddress,floatfix,nofootinbib,longbibliography]{revtex4-1}
\usepackage[utf8]{inputenc}
\usepackage{amsmath,amssymb,amsfonts,bm}
\usepackage{epsfig,epstopdf}
\usepackage{times}
\usepackage{graphics}
\usepackage{stmaryrd,latexsym}
\usepackage{color}
\usepackage[%dvipdfm,
colorlinks=true,
linkcolor=black,
breaklinks=true,
urlcolor=blue,
citecolor=blue]{hyperref}

\newcommand{\state}[1]{{\left|#1 \right\rangle} }

\newcommand{\X}{X(3872)}
\newcommand{\bd}{ \bm{  D  } }
\newcommand{\bdbar}{ \bm{  {\bar D } } }

\newcommand{\Amp}{\mathcal{A}}

\newcommand{\al}{&}

\begin{document}

\title{Revisiting $X(3872)\to D^0 \bar{D}^0 \pi^0$  in XEFT}

\author{Lin Dai}
\email{lin.dai@duke.edu}
\affiliation{Department of Physics,
         Duke University, Durham, NC 27705, USA}

\author{Feng-Kun Guo}
\email{fkguo@itp.ac.cn}
\affiliation{CAS Key Laboratory of Theoretical Physics, Institute of Theoretical Physics, Chinese Academy of Sciences, Beijing 100190, China}
\affiliation{School of Physical Sciences, University of Chinese Academy of Sciences, Beijing 100049, China}

\author{Thomas Mehen}
\email{mehen@phy.duke.edu}
\affiliation{Department of Physics,
         Duke University, Durham, NC 27705, USA}
%\date{\today}

\begin{abstract}
  The calculation of the decay $X(3872)\to D^0 \bar{D}^0 \pi^0$ in
  effective field theory is revisited to include final state $\pi^0 D^0$,
  $\pi^0 \bar{D}^0$and $D^0\bar{D}^0$ rescattering diagrams. These introduce significant
  uncertainty into the prediction for the partial width as a function of 
  the binding energy. The differential distribution in the pion energy is 
  also studied for the first time. The normalization of the distribution is again quite uncertain due to higher order effects but the shape of 
  the distribution is unaffected by higher order corrections. Furthermore 
  the shape of the distribution and the location of the peak are sensitive to the binding energy of $X(3872)$. The shape is strongly impacted by the presence of virtual $D^{*0}$ graphs which highlights the molecular nature of the $X(3872)$.
  Measurement of the pion energy distribution in the decay $X(3872)\to D^0
  \bar{D}^0 \pi^0$ can reveal interesting information about the binding
  nature of the $X(3872)$. 
  
\end{abstract}

\maketitle

\section{Introduction}

The $X(3872)$~\cite{Abe:2003hq} is the first of many exotic charmonium and bottomonium states found in various high energy experiments since 2003.  For reviews of these so-called $XYZ$ mesons and related exotic states, see Refs.~\cite{
Chen:2016qju,Richard:2016eis,Esposito:2016noz,Hosaka:2016pey,Lebed:2016hpi,Guo:2017jvc,Ali:2017jda,Olsen:2017bmm,Kou:2018nap,Cerri:2018ypt,Liu:2019zoy,Brambilla:2019esw}, and for a review dedicated to the $X(3872)$, see Ref.~\cite{Kalashnikova:2018vkv}. The quantum numbers of the $X(3872)$ have been determined to be $J^{PC}=1^{++}$\cite{Aaij:2013zoa}, which is consistent with it being a $\chi_{c1}$ state (and it is thus called $\chi_{c1}(3872)$ in the Review of Particle Physics (RPP)~\cite{Tanabashi:2018oca}, which reflects its quantum numbers without reference to the internal structure). A striking fact about the $X(3872)$ is that it is essentially degenerate with the $D^{*0}\bar{D}^0$ threshold. Currently, the particle data group has the $X(3872)$ mass being degenerate with $m_{D^{*0}}+m_{D^0}$ with an uncertainty of about 0.2 MeV. Therefore, the $X(3872)$ couples strongly to this channel and its wavefunction should contain a significant component of weakly bound $D^{*0}\bar{D}^0$ (plus the charge conjugate channel, which will be implied in what follows). Because of this it is widely believed that the $X(3872)$ is a hadronic molecule composed of $D^{*0}\bar{D}^0$.

If the $X(3872)$ is a weakly coupled bound state of $D^{*0}\bar{D}^0$ then ideas of effective range theory (ERT) can be applied to its decays. Voloshin used ERT to compute the decays $X(3872)\to D^0\bar{D}^0\pi^0$ and $X(3872)\to D^0\bar{D}^0\gamma$~\cite{Voloshin:2003nt}. For the former decay, the partial width for $X(3872)$ was predicted as a function of the binding energy, which was much more uncertain at the time of that publication. In the limit of zero binding energy the $D^{*0}$ in the $X(3872)$ is at rest and the partial width for $X(3872)\to D^0\bar{D}^0\pi^0$ is equal to that for $\Gamma[D^{*0}\to D^0 \pi^0] = 36.4\pm 0.98$ keV. \footnote{This value was obtained in Ref.~\cite{Mehen:2015efa} using the current measured values for $\Gamma[D^{*+}]$ and the branching fraction for $D^{*+}\to D^+ \pi^0$, and using isospin to relate $\Gamma[D^{*+}\to D^+ \pi^0]$ to
$\Gamma[D^{*0}\to D^0 \pi^0]$.}   For the current constraints on the binding energy the $X(3872)$ should be very close to this limit. However, it is difficult to test this prediction as the total width is only weakly constrained, $\Gamma[X(3872)] < 1.2$ MeV, and the branching fraction to $D^0 \bar{D}^0 \pi^0$ only is constrained to be greater than 40\% in the RPP~\cite{Tanabashi:2018oca}. 
\footnote{{A recent determination of the branching fraction to the $D^0\bar D^{*0}\pi^0$ mode gives $52.4^{+25.3}_{-14.3}\%$~\cite{Li:2019kpj}. It was criticized in Ref.~\cite{Braaten:2019ags} as the $X(3872)$ bound state feature was not taken into account. The latter reference also gives a branching fraction from the $X(3872)$ resonance feature for the $D^0\bar D^0\pi^0$ extracted from the $B$ decays as $(49\pm26)\%$.}}

ERT can be systematically improved upon using effective field theory (EFT). For the $X(3872)$ the relevant effective theory is XEFT~\cite{Fleming:2007rp}. In this theory the relevant degrees of freedom are the $D^0$, $\bar{D}^0$, $D^{*0}$, $\bar{D}^{*0}$, and $\pi^0$, all treated non-relativistically. In Ref.~\cite{Fleming:2007rp}, XEFT was used to calculate the partial width for the decay  $X(3872)\to D^0\bar{D}^0\pi^0$. The ERT prediction for $X(3872)\to D^0\bar{D}^0\pi^0$ emerges as the leading order in XEFT, and corrections from pion loops, range corrections, and higher dimension operators in the effective Lagrangian can be treated systematically.  In XEFT  pions can be treated perturbatively, and Ref.~\cite{Fleming:2007rp} found that pion loops gave very small corrections to the decay. By varying the effective range and coefficients of other operators, Ref.~\cite{Fleming:2007rp} was able to estimate the uncertainty in the ERT prediction for the partial width as a function of binding energy. 
For other applications of XEFT see Refs.\cite{Fleming:2008yn,Fleming:2011xa,Mehen:2011ds,Margaryan:2013tta,Braaten:2010mg,Canham:2009zq,Jansen:2013cba,Jansen:2015lha,Mehen:2015efa,Alhakami:2015uea,Braaten:2015tga}. For other EFT approaches to the $X(3872)$ see Refs.~\cite{AlFiky:2005jd,Baru:2011rs,Valderrama:2012jv,Nieves:2012tt,Baru:2013rta,Guo:2013nza,Guo:2013sya,Baru:2015nea,Schmidt:2018vvl}.

In Ref.~\cite{Guo:2017jvc}, it was pointed out that operators contributing to $\pi$-$D$ scattering lengths give rise to diagrams 
that are the same order in the power counting as those considered in Ref.~\cite{Fleming:2007rp}. The relevant
operators were first written down in Refs.~\cite{Guo:2008gp,Guo:2009ct}, which came after
Ref.~\cite{Fleming:2007rp}. The $\pi$-$D$ scattering lengths have been determined from recent lattice studies~\cite{Liu:2012zya,Mohler:2012na,Mohler:2012na,Guo:2018tjx}, and part of the goal of this paper is to
calculate the impact they have on the prediction for the partial
width of the $X(3872)$ in XEFT. In addition we also include
$D^0 \bar{D}^0$ rescattering effects, these were first studied in
Ref.~\cite{Guo:2014hqa}. We find that including these effects doubles the 
uncertainty in the prediction for the partial width found in Ref.~\cite{Fleming:2007rp}.  The theoretical uncertainty in the prediction for the partial width is rather large, for example, it is ${}^{+50}_{-30}\%$ for a binding energy of $0.2$ MeV. 

In this paper, we also calculate the differential spectrum, $d\Gamma[X(3872)\to D^0 \bar{D}^0 \pi^0]/dE_\pi$, where $E_\pi$ is the pion energy, for the first time. While the normalization of this rate suffers similar uncertainties, the shape of this distribution does not. Interestingly, this shape is sensitive to the binding energy and the molecular nature of the $X(3872)$. The distribution becomes more narrowly peaked near the maximal pion energy as the binding energy goes to zero. The location of the peak is insensitive to higher order corrections. The shape of the distribution is sensitive to the molecular nature of the $X(3872)$ because the $X(3872)$ couples to the final state through a virtual $D^{*0}$ propagator which is largest when the momentum of either $D$
meson in the final state goes to zero. The propagator has a pole at
$p_D^2=-\gamma^2$ and $p_{\bar{D}}^2=-\gamma^2$, where $p_D$,
$p_{\bar{D}}$ and $\gamma$ are the $D$ meson momentum, $\bar{D}$ meson momentum, and $X(3872)$ binding momentum, respectively. Thus, effect becomes more dramatic as the binding energy goes to zero. Amplitudes without the $D^*$ pole lead to broad $E_\pi$ distributions peaked at lower energy.

In the next section we briefly review XEFT as is relevant for our analysis and give the NLO amplitudes explicitly. In Section \ref{sec:numerical} we show the results of the calculations of the partial width and and the differential distribution in $E_\pi$. In Section \ref{sec:conclusions} we give our conclusions.

\section{NLO Partial Decay Amplitudes of $X(3872)$}

%-----------------------------------------
\begin{figure}[tb]
  \begin{center}
   \includegraphics[width=0.6\linewidth]{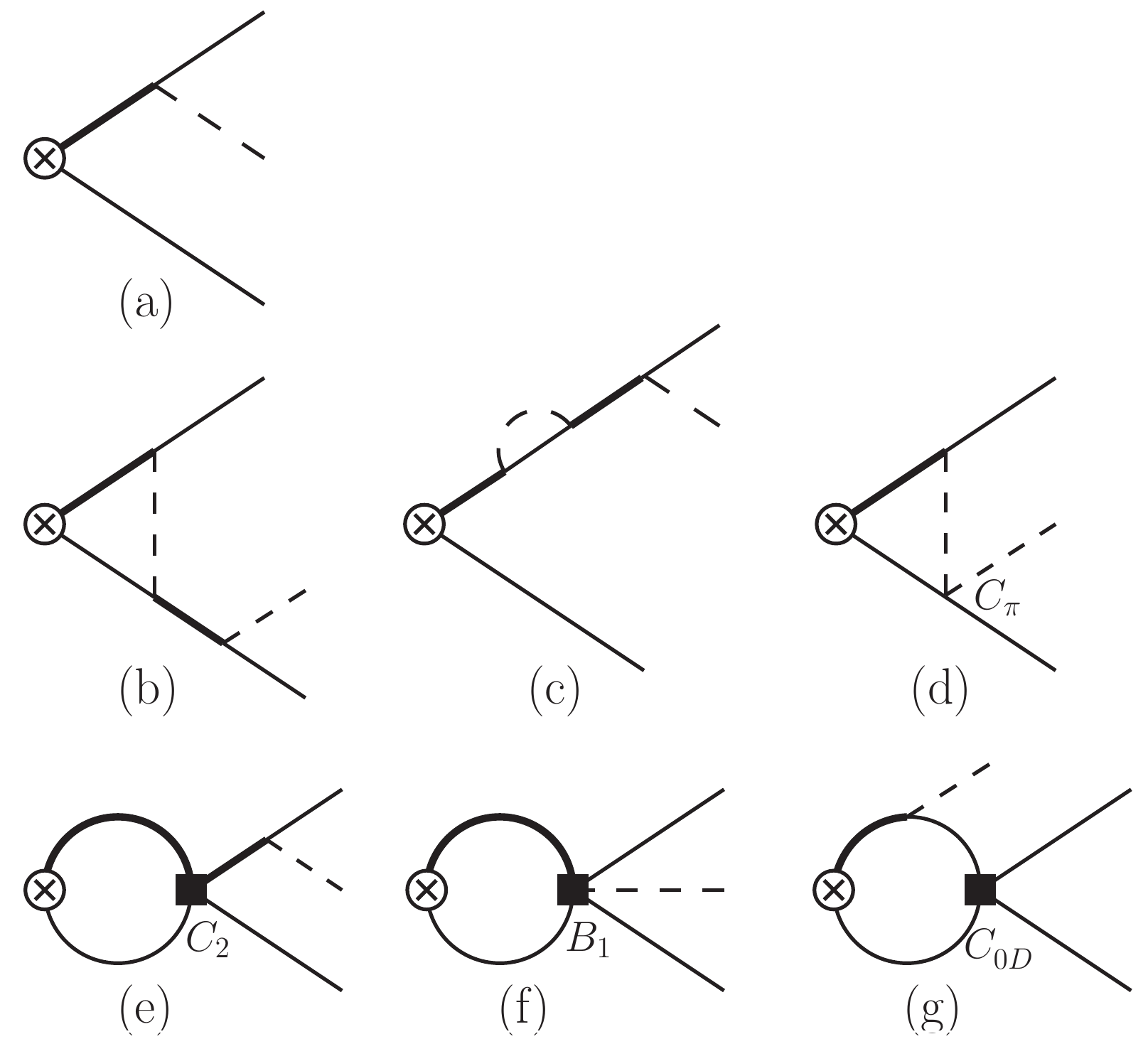}\\
   \caption{Feynman diagrams (reprinted from Refs.~\cite{Guo:2017jvc, Fleming:2007rp}) up to NLO for calculating the partial decay width of 
$\X\to D^0\bar D^0\pi^0$. The circled cross is the $\X$ field insertion. The thin solid, thick solid, and dashed lines represent $D^0(\bar{D}^0)$, $D^{*0}(\bar{D}^{*0})$, and $\pi^0$, respectively.
   \label{fig:XDDpi}}
  \end{center}
\end{figure}
%-----------------------------------------

XEFT~\cite{Fleming:2007rp} combines heavy hadron chiral perturbation theory~\cite{Wise:1992hn,Burdman:1992gh,Yan:1992gz} with the
Kaplan--Savage--Wise approach to describing low energy nucleon-nucleon
interactions~\cite{Kaplan:1998tg,Kaplan:1998we}.  Similar effective theories  have been developed for other
possible hadronic molecules which are located very close to thresholds, for example,  $Z_b(10610)$ and
$Z_b(10650)$~\cite{Mehen:2011yh,Mehen:2013mva} and the
$Z_c(3900)$~\cite{Wilbring:2013cha}. For a more detailed discussion of XEFT we refer the readers to the original work in Ref. \cite{Fleming:2007rp}. Here we simply quote the XEFT Lagrangian and comment on the power counting of Feynman diagrams relevant for our present analysis. 

The XEFT Lagrangian we use to calculate the partial decay rate of $X(3872)$ $\to$ $D^0\bar{D}^0\pi^0$ is \cite{Fleming:2007rp,Guo:2017jvc}
\begin{eqnarray}
 {\cal L} 
  &=& 
 \sum_{\bm{\phi}=\bd,\bdbar } \bm{\phi}^{\dagger} \bigg(i\partial_0 + 
\frac{\bm{\nabla}^2}{2 M_{D^{*0}}
		    }\bigg)\bm{\phi} 
 +  \sum_{{\phi}=D,\bar D } \phi^\dagger \bigg(i\partial_0 + 
\frac{\bm{\nabla}^2}{2 M_{D^0} } \bigg) \phi 
 + \pi^\dagger \bigg(i\partial_0 + \frac{\bm{\nabla}^2}{2 M_{\pi^0}}
   + \delta\bigg) \pi
%%%
\nonumber \\
&&+ \left[  
\frac{\bar{g}}{F_\pi} \frac{1}{\sqrt{ 2M_{\pi^0} } }
 \left( D \bd^\dagger \cdot \bm{\nabla}\pi  
   + \bar D^\dagger \bdbar \cdot \bm{\nabla}\pi^\dagger \right) + {\rm
 h.c.} \right]
\nonumber \\
 && 
- \,  
\frac{C_0}{2} \, \left(\bdbar D + \bd \bar D \right)^\dagger 
\cdot \left(\bdbar D + \bd \bar D \right) \nonumber\\
&&
+  \left[   \frac{C_2 }{16} \, 
\left(\bdbar D + \bd \bar D \right)^\dagger 
\cdot \left(\bdbar (\overleftrightarrow \nabla)^2 D 
        + \bd (\overleftrightarrow \nabla)^2 \bar D \right) + {\rm h.c.} \right]
\nonumber \\
%%%
&&+ \left[  \frac{B_1 }{\sqrt{2}}\frac{1}{\sqrt{2 M_{\pi^0}}} \left(\bdbar D + 
\bd \bar D \right)^\dagger \cdot D \bar{D} \bm{\nabla} \pi + {\rm 
h.c.}\right] \nonumber\\
%% new terms %%
&& + \frac{C_\pi}{2 M_{\pi^0}} \left( D^\dagger \pi^\dagger D \pi + 
\bar D^\dagger 
\pi^\dagger \bar D \pi \right) + C_{0D} D\,^\dagger\bar{D}^\dagger D\bar D \,,
\label{eq:XEFTlag}
\end{eqnarray}
where $\delta=\Delta-M_{\pi^0}\simeq7.0$~MeV with $\Delta = M_{D^{*0}}-M_{D^0}$, the pion decay constant is $F_\pi=91.9$~MeV, the coupling between pions and charmed mesons $\bar{g}\simeq 0.27$ fixed from the updated $D^{*+}$ decay width~\cite{Tanabashi:2018oca},\footnote{Notice that  $\bar{g}$ is related to the usually used one $g$ by $\bar g=g/2$. }  and $\overleftrightarrow \nabla = \overleftarrow \nabla - \overrightarrow \nabla$. The last line of Eq. (\ref{eq:XEFTlag}) gives rise to $\pi^0 D^0 (\bar{D}^0)$ and $D^0 \bar{D}^0$ rescattering which were not considered in the original Ref.~\cite{Fleming:2007rp}. The Feynman diagrams up to NLO that are relevant to the partial decay rate of $X(3872)$ $\to$ $D^0\bar{D}^0\pi^0$ are shown in Fig. \ref{fig:XDDpi}. 

To justify the necessity of including the $\pi^0 D^0 (\bar{D}^0)$ and $D^0 \bar{D}^0$ rescattering diagrams in studying the partial decay width of $X(3872)$ $\to$ $D^0\bar{D}^0\pi^0$, we review the power counting of Feynman diagrams in Fig.~\ref{fig:XDDpi} that was already discussed in detail in Ref.~\cite{Fleming:2007rp} and Ref.~\cite{Guo:2017jvc}. The relevant momenta involved in the decay dynamics are $\{p_D, p_{D^*},p_{\pi},\gamma,\mu\}$ where $\gamma$ is the binding momentum, $\sqrt{2\mu_0 B_X}$, with
$B_X$ is the binding energy and $\mu=\sqrt{\Delta^2-M^2_{\pi^0}}$. These momenta are the same order, which we generically denote as $Q$ which is also the power counting parameter. For the diagrams in Fig. \ref{fig:XDDpi}, each pion vertex contributes a factor of $Q$, each propagator $Q^{-2}$, and each loop integral $Q^5$. It was argued in Ref. \cite{Fleming:2007rp} that $C_0$ scales as $Q^{-1}$ which is responsible for the formation of $X(3872)$ bound state, and that $C_2$ and $B_1$ both scale as $Q^{-2}$ and together can be parameterized in terms of the effective range in $D^{*0}\bar{D}^0$ (or $\bar{D}^{*0}{D}^0$) scattering. The power counting of $C_{\pi}$ and $C_{0D}$ was discussed in Ref. \cite{Guo:2017jvc}. The $C_{\pi}$ contact term can be obtained by matching to the Heavy Hadron Chiral Lagrangian~\cite{Burdman:1992gh,Wise:1992hn,Yan:1992gz,Guo:2008gp}. The NLO investigation of the $\pi^0 D^0 \to \pi^0 D^0$ scattering indicates (using $D_s$ and $D$ mass splitting and lattice data) that the NLO amplitude ${\cal A}(\pi^0 D^0 \to \pi^0 D^0)$ is of order one, i.e., $C_{\pi} \sim Q^0$~\cite{Guo:2008gp,Guo:2009ct}. If $C_{0D}$ is assumed to be of order $Q^0$ as a natural choice, then the NLO diagram from $D^0\bar{D}^0$ rescattering (diagram (g) in in Fig.~\ref{fig:XDDpi}) scales as $Q^0$. On the other hand, if $D^0\bar{D}^0$ rescattering becomes non-perturbative, then $C_{0D}$ scales as $Q^{-1}$ which is similar to the scaling of $C_0$ discussed above and which gives even more significant contribution to the decay of $X(3872)$. With these basic scaling rules in hand, it is straightforward to check, in Fig. \ref{fig:XDDpi}, that LO diagram (a) scales as $Q^{-1}$ and NLO diagrams (b)--(f) scale as $Q^0$. In addition, depending on whether $C_{0D}$ scales as $Q^{0}$ or $Q^{-1}$, diagram (g) scales as $Q^0$ or $Q^{-1}$.

Next, we list all the decay amplitudes of $X(3872)$ up to NLO, all of which  are derived in the rest frame of the $X(3872)$. The LO amplitude from the tree-level diagram (a) in Fig.~\ref{fig:XDDpi} is
\begin{eqnarray}
  i\Amp_\text{LO} \al=\al \frac{\bar g \, \mu_0}{F_\pi \sqrt{M_{\pi^0}}} \vec{p}_\pi \cdot \vec{\epsilon}_X \left(\frac{1}{\vec p_D^{\,2}+\gamma^2} + \frac{1}{\vec p_{\bar D}^{\,2}+\gamma^2}   \right),
  \label{eq:LO-amplitude}
\end{eqnarray}
where $\mu_0$ is the reduced mass of the neutral $D$ and $D^*$ pair, $\vec p_{D(\bar D)}$ is the 3-momentum of the external $D^0(\bar D^0)$, $\vec p_\pi$ is the momentum of the final state $\pi^0$, $\vec{\epsilon}_X$ is the polarization vector of the $\X$, and $\gamma$ is the binding momentum.

The amplitude for diagram (b) in Fig.~\ref{fig:XDDpi} is given by
\begin{eqnarray}
  i\Amp_\text{(b)} \al=\al - \frac{\bar g^3 }{4 F_\pi^3 M_{\pi^0}^{3/2} }  \bigg\{ \frac{2\mu_0}{\vec p_D^{\,2}+\gamma^2} 
  \Big[ \vec{p}_D^{\,2} \vec{p}_\pi\cdot \vec{\epsilon}_X I_{1}^{(2)}(p_D) + \vec{p}_{D}\cdot \vec{p}_\pi \vec{p}_{D}\cdot \vec{\epsilon}_X \Big( I(p_{D}) - 2 I^{(1)}(p_{D}) + I_{0}^{(2)}(p_{D}) \Big) \Big] \nonumber\\
  \al\al +
  \frac{2\mu_0}{\vec p_{\bar D}^{\,2}+\gamma^2} 
  \Big[ \vec{p}_{\bar D}^{\,2} \vec{p}_\pi\cdot \vec{\epsilon}_X I_{1}^{(2)}(p_{\bar D}) + \vec{p}_{\bar D}\cdot \vec{p}_\pi \vec{p}_{\bar D}\cdot \vec{\epsilon}_X \Big( I(p_{\bar D}) - 2 I^{(1)}(p_{\bar D}) + I_{0}^{(2)}(p_{\bar D}) \Big) \Big] \bigg\} \,, 
  \label{eq:ampb}
\end{eqnarray}
 The loop integrals $I(p)$, $I^{(1)}(p)$, and $I^{(2)}_{0,1}(p)$ are defined in Appendix~\ref{app:loop}, and $m_1$, $m_2$ and $m_3$ should be replaced by the mass of $D^{*0}$, $D^0$ and $\pi^0$, respectively.
 
 Diagram (c) in Fig.~\ref{fig:XDDpi} does not contribute to the NLO calculation of the decay of $X(3872)\to D^0 \bar{D}^0 \pi^0$ \cite{Fleming:2007rp}. This is because the $D^{*0}$ and $\bar{D}^{*0}$ residual mass counterterms cancel the real part of diagram (c) while the imaginary part does not contribute at NLO since the LO diagram (a) is purely real. 

The amplitude for diagram (d) in Fig.~\ref{fig:XDDpi} is given by
\begin{eqnarray}
  i\Amp_\text{(d)} = \frac{\bar g C_\pi}{4 F_\pi M_{\pi^0}^{3/2} } \Big\{ \vec{p}_D\cdot \vec{\epsilon}_X \left[ I(p_D) - I^{(1)}(p_D) \right] + \vec{p}_{\bar{D}}\cdot \vec{\epsilon}_X \left[ I(p_{\bar D}) - I^{(1)}(p_{\bar D}) \right] \Big\}.
  \label{eq:ampd}
\end{eqnarray}
Again the loop integrals $I(p)$ and $I^{(1)}(p)$ are defined in Appendix~\ref{app:loop}, where $m_1$, $m_2$ and $m_3$ should be replaced by the mass of $D^{*0}$, $D^0$ and $\pi^0$, respectively.

The amplitude for diagram (e) reads
\begin{eqnarray}
  i \Amp_\text{(e)} = - C_2 \frac{ \bar g \mu_0^2}{4\pi F_\pi \sqrt{M_{\pi^0}}} \left(\Lambda_\text{PDS} - \gamma \right) \vec{p}_\pi \cdot \vec{\epsilon}_X  \left( \frac{\vec{p}_{\bar D}^{\,2} - \gamma^2}{\vec{p}_{\bar D}^{\,2} + \gamma^2} + \frac{\vec{p}_{D}^{\,2} - \gamma^2}{\vec{p}_{D}^{\,2} + \gamma^2} \right) ,
\end{eqnarray}
where $\Lambda_\text{PDS}$ is the scale introduced in the power divergence subtraction (PDS) scheme~\cite{Kaplan:1998tg} to regularize the ultraviolet (UV) divergence in the two-point bubble diagram, see Appendix~\ref{app:loop}.

The amplitude for diagram (f) reads
\begin{eqnarray}
  i \Amp_\text{(f)} = B_1 \frac{ \mu_0}{2\pi \sqrt{2M_{\pi^0}}} \left(\Lambda_\text{PDS} - \gamma \right) \vec{p}_\pi \cdot \vec{\epsilon}_X .
\end{eqnarray}
The amplitude for diagram (g) in Fig.~\ref{fig:XDDpi} is given by
\begin{eqnarray}
  i\Amp_\text{(g)} = \frac{ \bar g C_{0D}}{F_\pi \sqrt{M_{\pi^0}}} \vec{p}_\pi \cdot \vec{\epsilon}_X I(p_{\pi}) \, .
  \label{eq:ampg}
\end{eqnarray}
Here the loop integral $I(p)$ is as that defined in Appendix~\ref{app:loop}, where $m_1$, $m_2$ and $m_3$ should be replaced by the mass of $D^{*0}$, $D^0$ and $D^0$, respectively.

\section{Implications of $\pi^0 D^0$ and $D^0 \bar{D}^0$ rescattering}
\label{sec:numerical}

In this section, we numerically investigate the $\pi^0 D^0$ and $D^0 \bar{D}^0$ rescattering effects for the partial decay of $X(3872)\to D^0 \bar{D}^0 \pi^0$. The differential decay rate is
\begin{eqnarray}
\label{eq:decay-rate}
&&\frac{d \Gamma_{\mathrm{NLO}} }{d p_{D}^{2} d p_{\bar{D}}^{2}}=\frac{d \Gamma_{\mathrm{LO}} }{d p_{D}^{2} d p_{\bar{D}}^{2}}\left( 1+ \frac{\bar{g}^{2} \mu_0 \gamma}{3 \pi F_\pi^2} \left(\frac{4 \gamma^{2}-\mu^{2}}{4 \gamma^{2}+\mu^{2}}\right)+C_{2}\left(\Lambda_{\mathrm{PDS}}\right) \frac{\mu_{0} \gamma\left(\gamma-\Lambda_{\mathrm{PDS}}\right)^{2}}{\pi} \right) \nonumber\\
&&-\frac{\bar{g} \gamma}{8\sqrt{2} \pi^{3} F_{\pi}}\left(\sqrt{2}\frac{\bar{g} \mu_0}{F_{\pi}} C_{2}\left(\Lambda_{\mathrm{PDS}}\right)-B_{1}\left(\Lambda_{\mathrm{PDS}}\right)\right)\left(\Lambda_{\mathrm{PDS}}-\gamma\right)\left(\vec{p}_{\pi} \cdot \vec{\epsilon}_{X}\right)^{2}\left(\frac{1}{p_{D}^{2}+\gamma^{2}}+\frac{1}{p_{\bar{D}}^{2}+\gamma^{2}}\right) \nonumber\\
&&+ \frac{1}{8 \pi^{2}} \frac{\bar{g}^{4}}{F_{\pi}^{4}} \frac{\gamma}{M_{\pi^{0}}} \left[ \left(\vec{p}_{\pi} \cdot \vec{\epsilon}_{X}\right)^{2}\left(\frac{1}{p_{D}^{2}+\gamma^{2}}+\frac{1}{p_{\bar{D}}^{2}+\gamma^{2}}\right) \left( \frac{p_{D}^{2} {I}_{1}^{(2)}\left(p_{D}\right)}{P_{D}^{2}+\gamma^{2}}+ \frac{p_{\bar{D}}^{2} {I}_{1}^{(2)}\left(p_{\bar{D}}\right)}{P_{\bar{D}}^{2}+\gamma^{2}}\right) \right] \nonumber\\
 &&+\frac{1}{8 \pi^{2}} \frac{\bar{g}^{4}}{F_{\pi}^{4}} \frac{\gamma}{M_{\pi^{0}}} \left[ \vec{p}_{\pi} \cdot \vec{\epsilon}_{X}\left(\frac{1}{p_{D}^{2}+\gamma^{2}}+\frac{1}{p_{\bar{D}}^{2}+\gamma^{2}}\right) \left(\frac{\vec{p}_D \cdot \vec{\epsilon}_X \vec{p}_D \cdot \vec{p}_\pi }{p_{D}^{2}+\gamma^{2}}\left( I(p_D)-2I^{(1)}(p_D)+I^{(2)}_0(p_D)\right)\right. \right. \nonumber\\
 &&+\left. \left.  \frac{\vec{p}_{\bar{D}} \cdot \vec{\epsilon}_X \vec{p}_{\bar{D}} \cdot \vec{p}_\pi }{p_{\bar{D}}^{2}+\gamma^{2}} \left( I(p_{\bar{D}})-2I^{(1)}(p_{\bar{D}})+I^{(2)}_0(p_{\bar{D}})\right) \right)\right] \nonumber\\
 &&+ \frac{C_{\pi} \bar{g}^{2} \gamma\,\vec{p}_{\pi} \cdot \vec{\epsilon}_{X}}{16 \pi^{2} F_{\pi}^{2} M_{\pi^0} \mu_{0}} \left(\frac{1}{p_{D}^{2}+\gamma^{2}}+\frac{1}{p_{\bar{D}}^{2}+\gamma^{2}}\right) \left[ \vec{p}_{D}\cdot \vec{\epsilon}_X\left(I^{(1)}(p_{D})-I(p_{D}) \right)+\vec{p}_{\bar{D}}\cdot \vec{\epsilon}_X\left(I^{(1)}(p_{\bar{D}})-I(p_{\bar{D}}) \right) \right]\nonumber \\
 &&+\frac{C_{0 D} \bar{g}^2 \gamma}{4 \pi^{2} F_\pi^2 \mu_0}\left(\vec{p}_{\pi} \cdot \vec{\epsilon}_{X}\right)^{2} I(p_\pi)\left(\frac{1}{p_{D}^{2}+\gamma^{2}}+\frac{1}{p_{\bar{D}}^{2}+\gamma^{2}}\right).
 \end{eqnarray}
where the last two lines come from the contributions of the two new interactions (i.e., $\pi^0 D^0$ and $D^0\bar{D}^0$ contact terms) in the Lagrangian. We have numerically checked that if the two new interaction terms ($\pi^0 D^0 $ and $D^0\bar{D}^0$ scattering) are ignored, Eq. (\ref{eq:decay-rate}) agrees with the corresponding expression in Ref. \cite{Fleming:2007rp}.

\begin{figure}[htbp]
\begin{center}
\includegraphics[width=14cm]{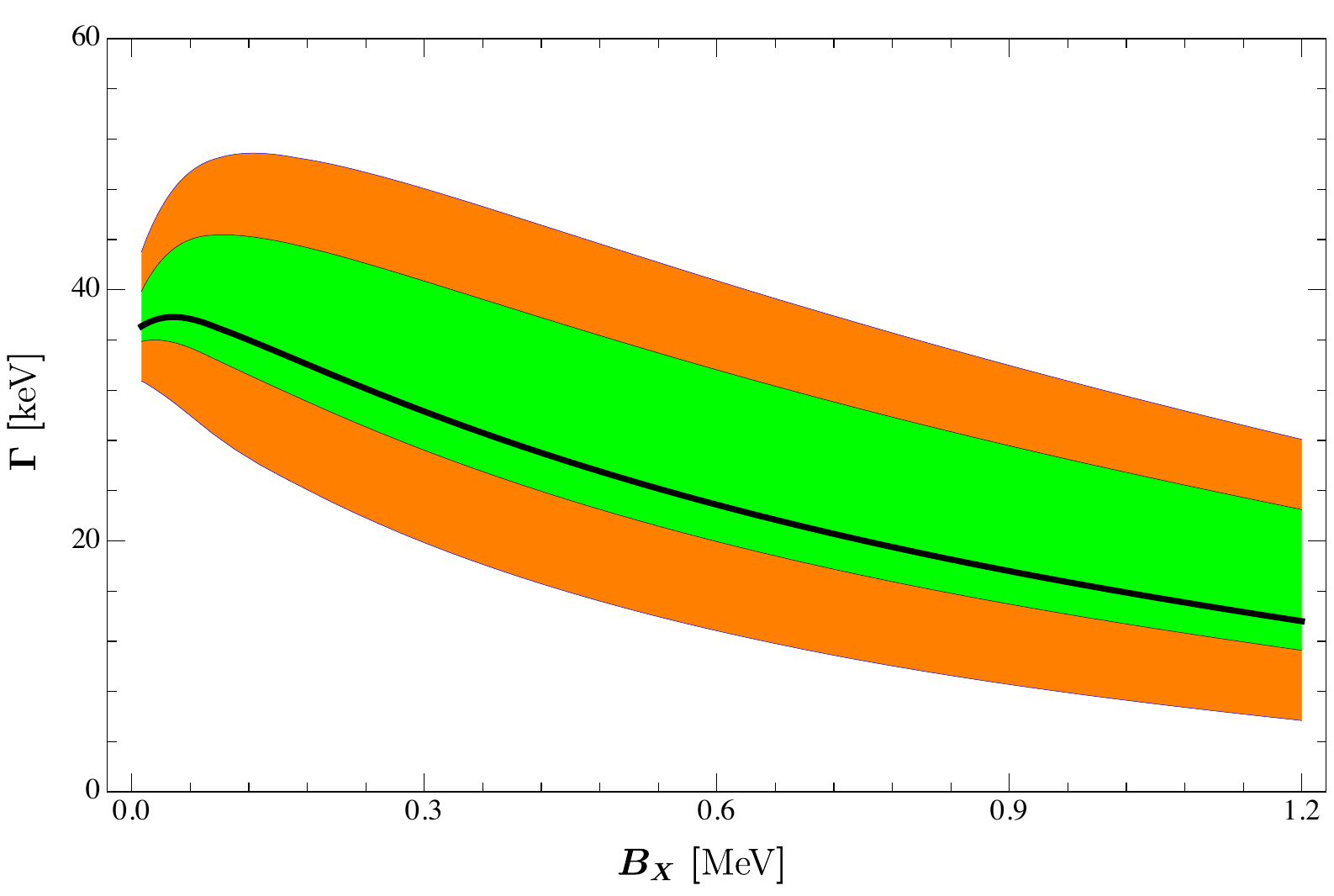}
\caption{Partial decay rate of $X(3872)\to D^0\bar{D}^0\pi^0$ versus the binding energy $B_X$ of $X(3872)$. The central black solid line is the LO decay rate. The green band is the NLO correction that has already been obtained in Ref.~\cite{Fleming:2007rp}. The orange band is the NLO correction coming from the two new interactions, i.e., $\pi^0 D^0 $ and $D^0\bar{D}^0$ rescattering.}
\label{fig:Gamma-BX}
\end{center}
\end{figure}

In Fig. \ref{fig:Gamma-BX}, we show NLO corrections to the partial decay rate of $X(3872)\to D^0\bar{D}^0\pi^0$.  The green band reproduces the results already obtained in Ref. \cite{Fleming:2007rp}, where $\pi^0 D^0 $ and $D^0\bar{D}^0$ rescattering were not included. The orange band is the additional correction due to the $\pi^0 D^0 $ and $D^0\bar{D}^0$ rescattering terms in the Lagrangian. In Fig. \ref{fig:Gamma-BX}, $C_{\pi}$ is set to be $4.1 \pm 0.7$ GeV$^{-1}$ which is derived in Appendix~\ref{sec:scatt-length}. {Currently, little is known about $C_{0D}$. We choose to vary it between a natural range of $[-1, 1]~{\rm fm}^2$ to estimate its impact on the results.} In particular $C_{0D} = 1~{\rm fm}^2$ corresponds to having a $D^0\bar{D}^0$ bound state near threshold with scattering length approximately equal to $a$ $\sim$ $-\frac{m_{D}C_{0D}}{4\pi}=-\frac{1}{262} ~{\rm MeV}^{-1}$ \cite{Guo:2014hqa}. All other parameters are set to be the same as those in Ref.~\cite{Fleming:2007rp}. 
Note that $\bar{g}$   differs slightly from the value used in Ref.~\cite{Fleming:2007rp} to be consistent with the most recent measurement of the decay width for $\Gamma[D^{*+}]$~\cite{Tanabashi:2018oca}, which is the reason why the central line of Fig.~\ref{fig:Gamma-BX} is slightly lower than that in Ref.~\cite{Fleming:2007rp}). The two new interactions give rise to a correction up to about $25\%$. Interestingly, $D^0\bar{D}^0$ rescattering contribution dominates that of the $\pi^0 D^0$ rescattering whose correction to the LO decay rate is less than $1\%$. This is reminiscent of the results obtained in Ref.~\cite{Fleming:2007rp}, where it is found that the contact interaction contribution dominates that from the pion exchange diagrams.

\begin{figure}[htbp]
\begin{center}
\includegraphics[width=15cm]{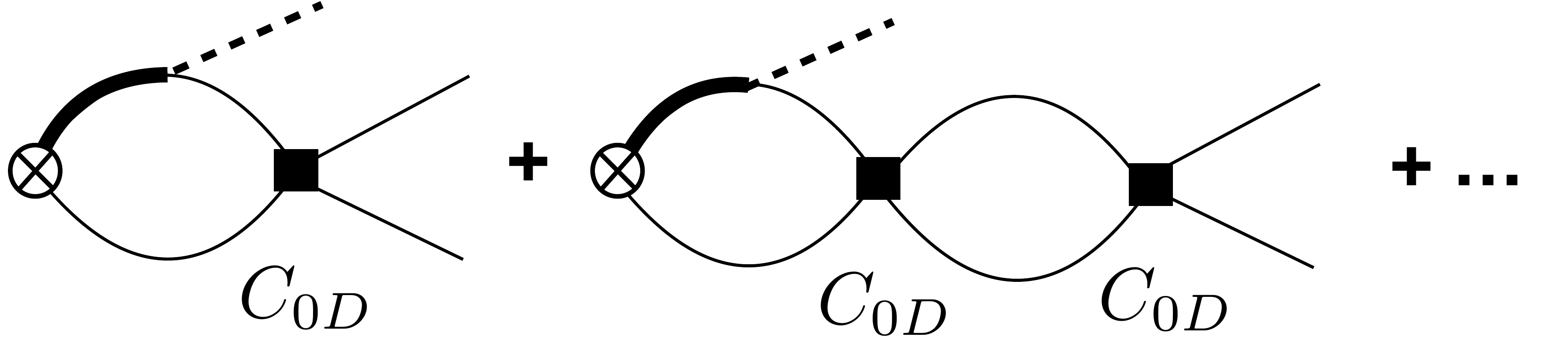}
\caption{Resumming the $D^0\bar{D}^0$ rescattering diagram.}
\label{fig:FSI-sum}
\end{center}
\end{figure}

\begin{figure}[htbp]
\begin{center}
\includegraphics[width=10cm]{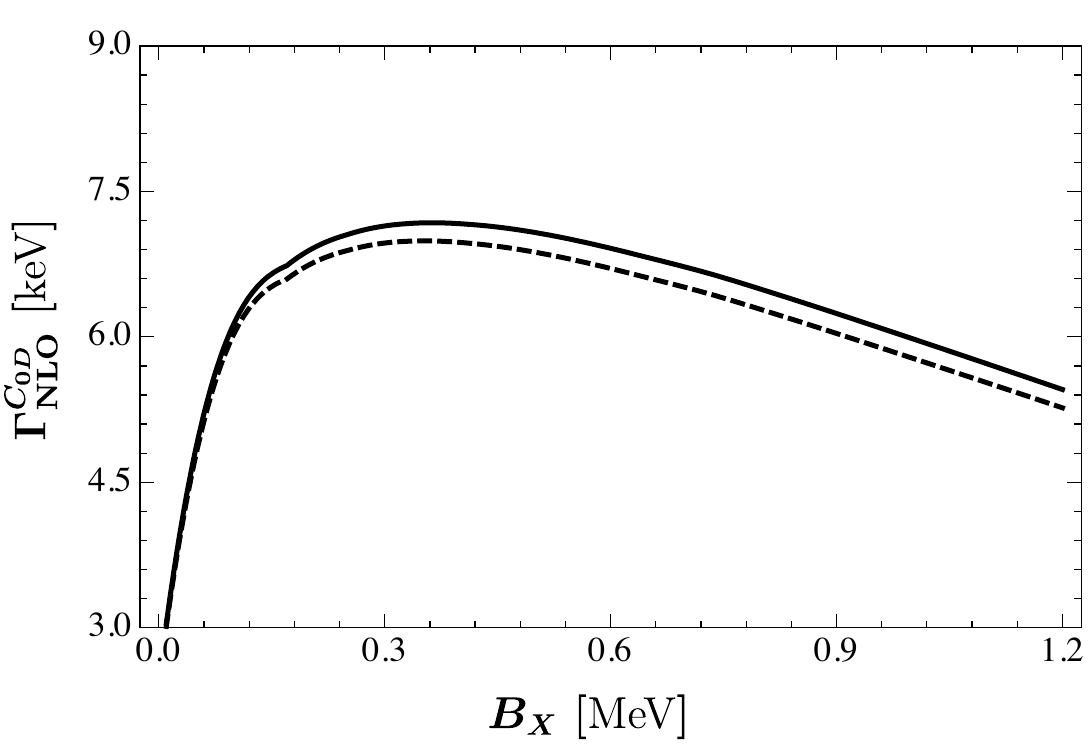}
\caption{The NLO corrections from the $D^0\bar{D}^0$ rescattering diagram. Solid and dashed lines correspond to contributions from the $D^0\bar{D}^0$ rescattering diagrams without and with resummation shown in Fig.~\ref{fig:FSI-sum}, respectively. }
\label{fig:CDNLO}
\end{center}
\end{figure}

Since the $D^0\bar{D}^0$ rescattering gives a large NLO correction to the partial decay rate of $X(3872)\to D^0\bar{D}^0\pi^0$, it is interesting to see the resummation effect of final state rescattering as shown in Fig. \ref{fig:FSI-sum}. The resummation is equivalent to replacing $C_{0D}$ with the effective range expansion \cite{Kaplan:1998we}
\begin{equation}
C_{0D} \to -\frac{4\pi}{m_D}\frac{1}{1/a+ip}
\end{equation}
where $a$ is the $D^0\bar{D}^0$ scattering length (set to be $-\frac{1}{262}$ ${\rm MeV}^{-1}$ as mentioned above) and $p=|(\vec{p}_D-\vec{p}_{\bar{D}})/2|$. The correction from such a resummation is not significant, though, as is shown in Fig. \ref{fig:CDNLO}, where the dashed line (from the resummed $D^0\bar{D}^0$ rescattering diagram) only gives a small modification to the solid line (from the $D^0\bar{D}^0$ rescattering diagram without resummation). Overall, the effect of resumming $D^0\bar{D}^0$ rescattering diagram shown in Fig. \ref{fig:FSI-sum} contributes to less than $1\%$ correction to the LO partial decay rate of $X(3872)\to D^0\bar{D}^0\pi^0$.

Next we investigate the $\pi^0$ kinetic energy, $E_\pi$, distribution from $X(3872)$ decaying to $D^0\bar{D}^0\pi^0$. The analytic expression can be obtained readily from Eq. (\ref{eq:decay-rate}) by changing variable from $p_{\bar{D}}$  to $E_\pi$ using energy conservation in the $X(3872)$ rest frame
\begin{equation}
p_{\bar{D}}^{2}=2m_{D}\left(M_X-m_{\pi}-2m_{D}-E_\pi-\frac{p_{D}^{2}}{2 m_{D}}\right),
\label{eq:pdbar}
\end{equation}
which leads to
\begin{equation}
\frac{d \Gamma}{d E_{\pi} d p_{D}^{2}}=2m_D \frac{d \Gamma}{d p_{D}^{2} d p_{\bar{D}}^{2}}.
\end{equation}
where the prefactor $2m_D$ on the right hand side comes from the Jacobian for changing variables, and $p_{\bar{D}}$ is expressed in terms of $E_{\pi}$ and $p_D$ using Eq. (\ref{eq:pdbar}). Then the $\pi^0$ kinetic energy $E_\pi$ distribution is obtained by integrating over $p_D$.

\begin{figure}[htbp]
\begin{center}
\includegraphics[width=18cm]{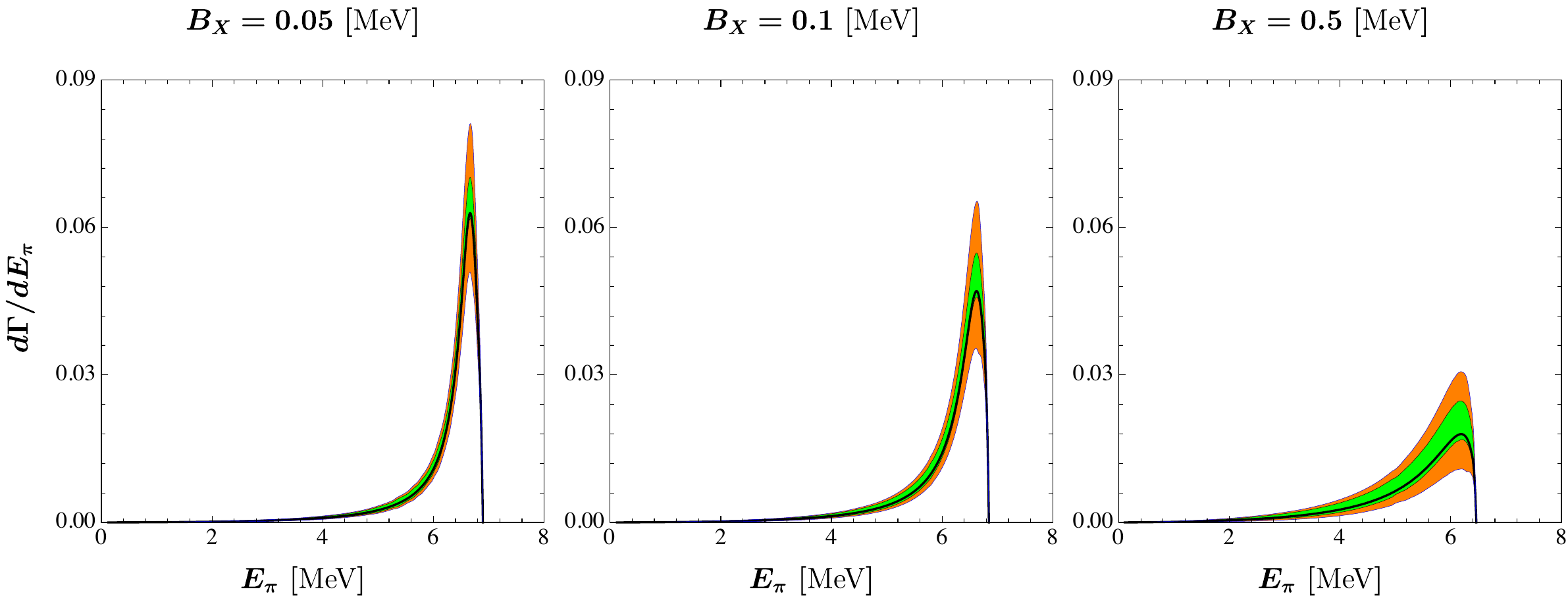}
\caption{The $\pi^0$ kinetic energy distribution for the decay process $X(3872)\to D^0\bar{D}^0\pi^0$. Color and line conventions for the plots are the same as those used in Fig.~\ref{fig:Gamma-BX}.}
\label{fig:Gamma-Epi}
\end{center}
\end{figure}

\begin{figure}[htbp]
\begin{center}
\includegraphics[width=18cm]{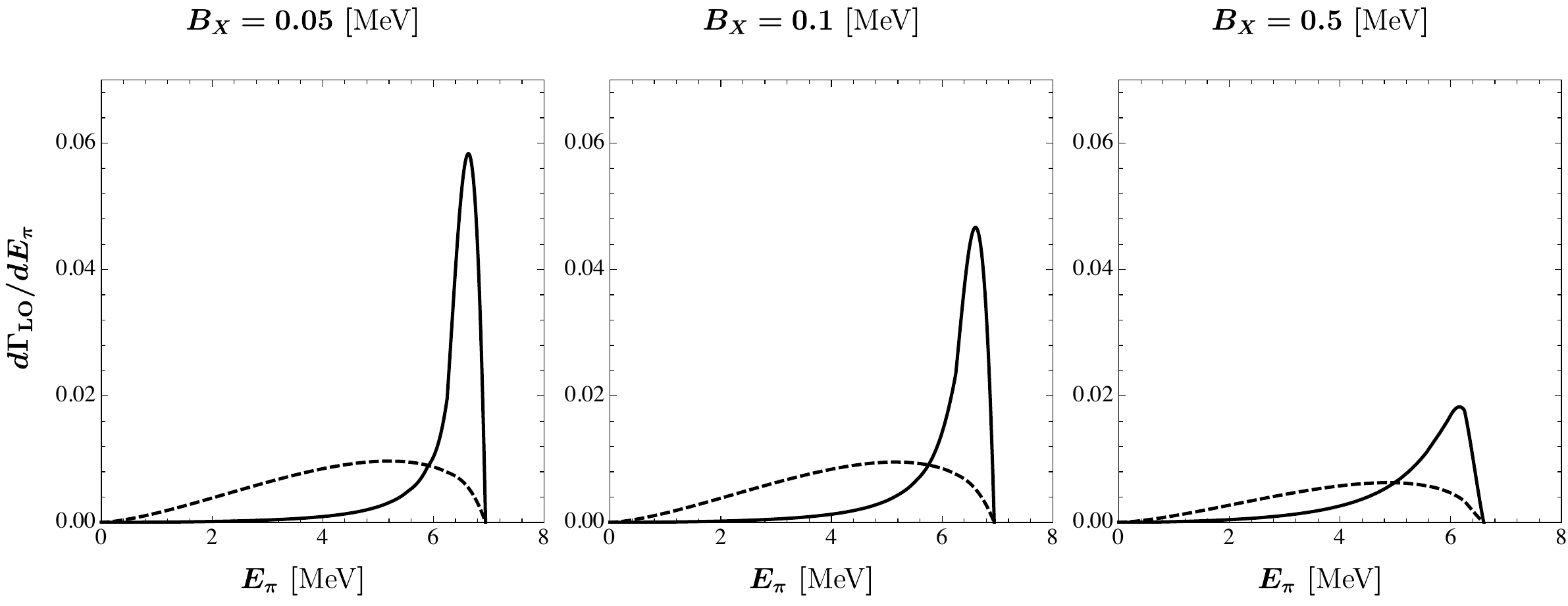}
\caption{Effects of the dynamics of $D^{0*}$ (and $\bar{D}^{0*}$) as indicated in the LO amplitude Eq.~(\ref{eq:LO-amplitude}). The dashed curves are normalized to have the same areas as the solid curves in each panel respectively. The solid curves are the LO $E_{\pi}$ distribution from Eq.~(\ref{eq:LO-amplitude}) and dashed curves are the LO $E_{\pi}$ distribution from Eq.~(\ref{eq:LO-amplitude}) with the $D^{0*}$ (and $\bar{D}^{0*}$) propagators set to a constant.}
\label{fig:pDD}
\end{center}
\end{figure}

\begin{figure}[htbp]
\begin{center}
\includegraphics[width=18cm]{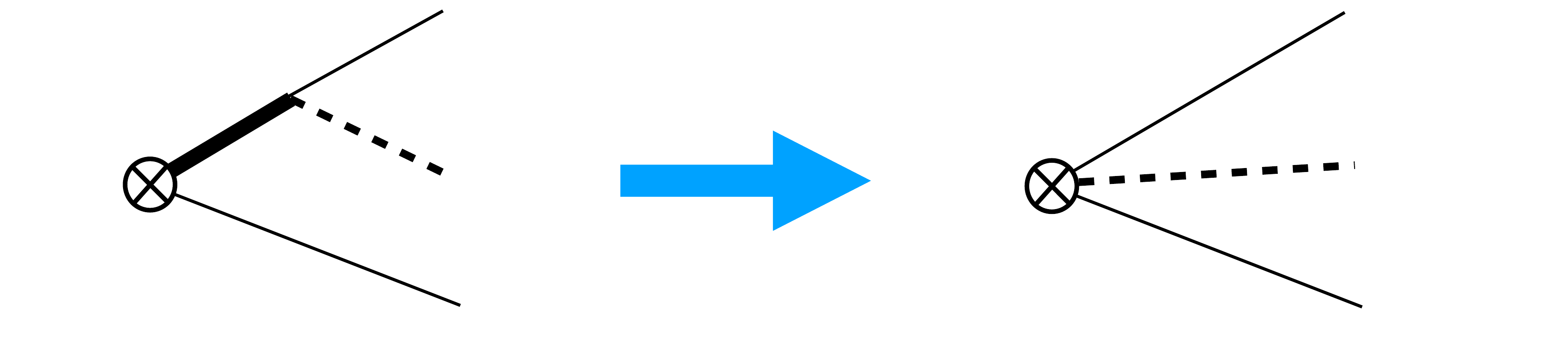}
\caption{Shrinking of the $D^{0*}$ ($\bar{D}^{0*}$) propagators to contact interactions with $X(3872)$.}
\label{fig:pShrink}
\end{center}
\end{figure}

Figure~\ref{fig:Gamma-Epi} shows the $\pi^0$ kinetic energy, $E_{\pi}$, distributions for different binding energies, $B_X$, of $X(3872)$. As $B_X$ decreases, the location of the peaks shifts to higher energies and the peak is higher and more narrow.  As shown in Fig.~\ref{fig:Gamma-BX}, the $D^0\bar{D}^0$ rescattering gives rise to large NLO corrections, which makes the extraction of $X(3872)$ binding energy from the partial decay rate difficult without further knowledge of the $D\bar D$ interaction (see also Ref.~\cite{Guo:2014hqa}). Figure~\ref{fig:Gamma-Epi} shows, however, the location of the peak of the $E_\pi$ distribution is insensitive to NLO corrections, so it could be a better observable for extracting properties of $X(3872)$. While the three-body phase-space is important in determining the overall features of $E_{\pi}$ distributions, the sharpness of the peaks is due to the pole from the virtual $D^{0*}$ (and $\bar{D}^{0*}$) propagator,  which is a consequence of the molecular nature of the $X(3872)$. To further illustrate this point, we do a simple LO analysis of $E_{\pi}$ distribution based on LO amplitude Eq. (\ref{eq:LO-amplitude}). In Fig. \ref{fig:pDD}, the solid curves are the same as the LO curves shown in Fig. \ref{fig:Gamma-Epi} and dashed curves are LO $E_{\pi}$ distribution obtained by setting the $D^{0*}$ (and $\bar{D}^{0*}$) propagators in Eq. (\ref{eq:LO-amplitude}) to a constant which effectively shrinks the propagators to contact interactions with $X(3872)$ as shown in Fig.~\ref{fig:pShrink}. Thus the dashed curves come from pure three-body phase-space ($\times p_\pi^2$), while for the solid curves the location and width of peaks have genuine information about the $D^{0*}$ (and $\bar{D}^{0*}$) binding in the $X(3872)$.

\section{Conclusions}
\label{sec:conclusions}

In this paper we revisited the XEFT calculation of $X(3872)\to D^0 \bar{D}^0 \pi^0$ first performed in Ref.~\cite{Fleming:2007rp}. We included corrections from the $\pi^0 D^0$, $\pi^0\bar{D}^0$ and $D^0 \bar{D}^0$ rescattering. The $D^0 \bar{D}^0$ rescattering is by far the most important. The total uncertainty in the partial width is of order $50\%$ and the $D^0 \bar{D}^0$ rescattering accounts for about half. 
We also calculate the differential distribution in the pion energy. This turns out to be quite interesting. Because of poles in diagrams with virtual $D^{*0}(\bar{D}^{*0})$ mesons, the distribution is highly peaked near maximal pion energy. The shape of this distribution is sensitive to the binding energy of the $X(3872)$. As the binding energy approaches zero, the location of the peak shifts upwards and the peak becomes narrower and higher. Unlike the partial width, the shape of the distribution and the location of the peak are insensitive to higher order corrections. It would be very interesting to make an experimental measurement of this distribution, in particular before the proposal of measuring the $X(3872)$ binding energy using triangle singularity~\cite{Guo:2019qcn} is realized experimentally. 

\acknowledgments T.M thanks the Institute of Theoretical Physics, Chinese Academy of Sciences (CAS), where he was Peng Huanwu Visiting Professor and this work was initiated. L.D. and T.M. are supported by in part by Director, Office of Science, Office of Nuclear Physics, of the U.S. Department of Energy under grant number DE-FG02-05ER41368. L.D. and T.M. are also supported in part by the U.S. Department of Energy, Office of Science, Office of Nuclear Physics, within the framework of the TMD Topical Collaboration. F.-K.G. is supported in part by the National Natural Science Foundation of China (NSFC) and  the Deutsche Forschungsgemeinschaft  through the funds provided to the Sino-German Collaborative Research Center  CRC110 ``Symmetries and the Emergence of Structure in QCD"  (NSFC Grant No. 11621131001), by the NSFC under Grants No. 11835015 and No.~11847612, by the CAS under Grant No. QYZDB-SSW-SYS013, and by the CAS Center for Excellence in Particle Physics (CCEPP).

\begin{appendix}

\section{Loop integrals}
\label{app:loop}
\renewcommand{\theequation}{\thesection.\arabic{equation}}
\setcounter{equation}{0}

Here we define the loop integrals involved in the calculations. We choose to work in the rest frame of the decay particle, $P^\mu = (M,\vec{0})$. The basic scalar 3-point loop integral is UV convergent, and can be worked out 
as~\cite{Guo:2010ak}\footnote{Note that the 3-point loop integrals defined here are related to those in Ref.~\cite{Guo:2010ak} by multiplying a factor of $1/(8 m_1 m_2 m_3)$.}
\begin{eqnarray}%
I(q) \al=\al i \int\!\frac{d^dl}{(2\pi)^d}
\frac{1}{ \left(l^0-\frac{\vec{ l}\ ^2}{2m_1}+i\epsilon\right)
\left( -b_{12} -l^0 - \frac{\vec{ l}\ ^2}{2m_2}+i\epsilon\right)
\left[l^0+b_{12}-b_{23}-\frac{(\vec{ l}-\vec{
q})^2}{2m_3}+i\epsilon\right] } \nonumber\\
\al=\al 4\mu_{12}\mu_{23}
\int\!\frac{d^{d-1}l}{(2\pi)^{d-1}} \frac{1}{\left(\vec{ l}\
^2+c_1-i\epsilon\right) \left(\vec{ l}\ ^2-\frac{2\mu_{23}}{m_3}\vec{
l}\cdot\vec{
q}+c_2-i\epsilon\right)} \nonumber\\
\al=\al 4\mu_{12}\mu_{23} \int_0^1dx
\int\!\frac{d^{d-1}l}{(2\pi)^{d-1}} \frac{1}{\left[\vec{ l}\
^2-ax^2+(c_2-c_1)x+c_1-i\epsilon\right]^2} \nonumber\\
\al=\al \frac{4\mu_{12}\mu_{23}}{(4\pi)^{(d-1)/2}} \Gamma\left( \frac{5-d}{2} \right) 
\int_0^1 dx  \left[ -a x^2 + (c_2 - c_1) x + c_1 - i \epsilon \right]^{(d-5)/2} \nonumber\\
\al=\al \frac{\mu_{12}\mu_{23}}{2\pi} \frac{1}{\sqrt{a}}
\left[ \tan^{-1}\left(\frac{c_2-c_1}{2\sqrt{ac_1}}\right) +
\tan^{-1}\left(\frac{2a+c_1-c_2}{2\sqrt{a(c_2-a)}}\right) \right],
\end{eqnarray}%
where $\mu_{ij}=m_im_j/(m_i+m_j)$ are the reduced masses, $b_{12} =
m_1+m_2-M$, $b_{23}=m_2+m_3+q^0-M$, and
\begin{eqnarray}%
a = \left(\frac{\mu_{23}}{m_3}\right)^2 \vec{ q}\ ^2, \quad c_1 =
2\mu_{12}b_{12}, \quad c_2=2\mu_{23}b_{23}+\frac{\mu_{23}}{m_3}\vec{
q}\ ^2.
\end{eqnarray}%
There is no pole for $d\leq4$, and we have taken $d=4$ in the last step.

We also need the vector and tensor loop integrals which are defined
as
\begin{equation}%
q^i I^{(1)}(q) \equiv i \int\!\frac{d^dl}{(2\pi)^d} l^i \times [\text{integrand of }I(q)],
\end{equation}%
and
\begin{eqnarray}%
\label{Aeq:tensorloop} 
\al\al q^iq^j I^{(2)}_0(q) + \delta^{ij}\vec{ q}\ ^2
I^{(2)}_1(q) \equiv i \int\!\frac{d^dl}{(2\pi)^d} {l^il^j}\times [\text{integrand of }I(q)].
\end{eqnarray}%
They can be expressed in terms of the scalar 2-point and 3-point loop integrals as
\begin{eqnarray}%
\label{Aeq:I1}
 I^{(1)}(q) \al=\al \frac{\mu_{23}}{a m_3}
\left[B(c_2-a)-B(c_1) +  \frac{1}{2} (c_2-c_1) I(q)\right],
\\ \label{Aeq:I20}%
\vec{ q}\ ^2 I^{(2)}_0(q) \al=\al \frac{d-3}{d-2} B(c_2-a) +
\frac{c_1}{d-2} I(q) + \frac{d-1}{d-2} (c_2-c_1) \frac{m_3}{2\mu_{23}}
I^{(1)}(q), \\ \label{Aeq:I21}%
\vec{ q}\ ^2 I^{(2)}_1(q) \al=\al \frac{1}{d-2} B(c_2-a) -
\frac{c_1}{d-2} I(q) - \frac{1}{d-2} (c_2-c_1) \frac{m_3}{2\mu_{23}}
I^{(1)}(q),
\end{eqnarray}%
where the function $B(c)=2\mu_{12}\mu_{23} \Sigma(c)$ with $\Sigma(c)$ defined as (in the PDS scheme)
\begin{eqnarray}%
\Sigma(c) \al\equiv\al \left( \frac{\Lambda_{\rm PDS}}{2} \right)^{4-d}\! 
\int\!\frac{d^{d-1}l}{(2\pi)^{d-1}} \frac{1}{\vec{
l}\ ^2+c-i\epsilon} \nonumber\\
\al=\al \left( \frac{\Lambda_{\rm PDS}}{2} \right)^{4-d} (4\pi)^{(1-d)/2}\Gamma\left(\frac{3-d}{2}\right)(c-i\epsilon)^{(d-3)/2}
\nonumber\\
  &=& \frac{1}{4\pi} \left(\Lambda_{\rm PDS} -\sqrt{c-i\epsilon} \right) .
\label{Aeq:B}
\end{eqnarray}%
{Notice that although the expression for $I^{(2)}_0(q)$ contains the $B$ function, it is in fact UV convergent since  $l^il^j$ in Eq.~\eqref{Aeq:tensorloop} is converted to $q^iq^j$, a factor of the external momentum, for this term. The $(d-3)$ factor in front of the $B$ function in Eq.~\eqref{Aeq:I20} cancels the $1/(d-3)$ pole term in the $B$ function. }

\section{${D\pi}$ scattering length}
\label{sec:scatt-length}

In this appendix, the numerical value of $C_{\pi}$ used in our plots is derived. We use the isospin phase convention:
\begin{equation}
    \state{\pi^+} = - \state{1,+1},\qquad
   \state{D^+} = - \state{\frac12,+\frac12} ,
\end{equation}
while all the other $D$ and pion fields take the positive sign. With this convention, the $D^0\pi^0$ is expressed in terms of isospin eigenstates as 
\begin{equation}
    \state{D^0\pi^0 } = \sqrt{\frac{2}{3}}\, \state{\frac32,-\frac12}  + \frac{1}{\sqrt{3}}\, \state{\frac12,-\frac12} .
\end{equation}
Then the $D^0\pi^0\to D^0\pi^0$ scattering length receives contributions from both isospin $I=1/2$ and $I=3/2$ channels as
\begin{equation}
    a_{D^0\pi^0} = \frac{2}{3} a_{D\pi}^{3/2} + \frac{1}{3} a_{D\pi}^{1/2} = (0.06\pm0.01)~\text{fm} ,
\end{equation}
where we have used $a_{D\pi}^{1/2}=0.37^{+0.03}_{-0.02}$~fm and $a_{D\pi}^{3/2}=-(0.100\pm0.002)$~fm from Ref.~\cite{Liu:2012zya}.

Then the dimensionless scattering amplitude at threshold with the relativistic normalization is
\begin{eqnarray}
  \Amp_{D^0\pi^0}(\sqrt{s}=M_{D^0}+M_{\pi^0}) = 8\pi (M_{D^0}+M_{\pi^0}) a_{D^0\pi^0} .
\end{eqnarray}
Matching the above expression to $2M_{D^0} 2 M_{\pi^0} C_\pi/(2 M_{\pi^0})$ gives
\begin{equation}
    C_\pi = 4\pi \left(1+ \frac{M_{\pi^0}}{M_{D^0}}\right) a_{D^0\pi^0}  = (4.1\pm0.7)\, {\rm GeV}^{-1}.
\end{equation}

\end{appendix}

\bibliography{xeft}

\end{document}